# AI-Enhanced Decision-Making for Sustainable Supply Chains: Reducing Carbon Footprints in the USA

**MD Rokibul Hasan**
Research Scholar, MBA in Business Analytics, Gannon University, Erie, PA

**Abstract**

Organizations increasingly need to reassess their supply chain strategies in the rapidly modernizing world towards sustainability. This is particularly true in the United States, where supply chains are very extensive and consume a large number of resources. This research paper discusses how AI can support decision-making for sustainable supply chains with a special focus on carbon footprints. These AI technologies, including machine learning, predictive analytics, and optimization algorithms, will enable companies to be more efficient, reduce emissions, and display regulatory and consumer demands for sustainability, among other aspects. The paper reviews challenges and opportunities regarding implementing AI-driven solutions to promote sustainable supply chain practices in the USA.

## Introduction

As per Hasan et al. (2024), supply chains play a critical role in the global economy by facilitating the flow of goods and products across borders. But they also contribute much to environmental degradation. According to the World Economic Forum, over 60 % of global carbon emissions are from supply chains, primarily driven by logistics, transportation, and manufacturing. According to the Environmental Protection Agency, an estimation in the United States shows that the transport sector is responsible for about 29% of the total GHG emissions in 2023. Chen et al. (2024), posited that as climate change grows in world consciousness, presses the need for carbon footprint reduction. Companies in the USA are increasingly feeling the strain to automate supply chains and minimize their environmental impact. Khan et al.(2024), asserted that Recently, the regulatory framework known as the Inflation Reduction Act of 2022 has become another compelling reason for organizations to transition into greener operations. Companies will be circumventing pressures by resorting to sophisticated technologies using AI that will optimize their supply chains in ways intended to reduce their environmental impact.

According to Buiya et al. (2024), Artificial Intelligence (AI) has emerged as a revolutionary technology capable of enhancing decision-making across different domains, encompassing supply chain management. AI for sustainable supply chains undoubtedly helps organizations reduce their emission levels, utilize their resources more efficiently, and reach their environmental objectives. The subsequent research paper explores how AI can improve decision-making in supply chains, with special emphasis on reducing carbon footprint in the USA. The report encompasses challenges, AI applications, case studies, and the outlook of AI promoting sustainability.

## The Need for Sustainability in Supply Chains:
### Supply Chain and Environmental Impact

Islam et al. (2023a), argued that in a highly interconnected and resource-restrained world, the need for sustainable supply chains has never been more pressing. Traditional approaches to supply chain management are mostly directed toward efficiency and cost reduction but at great environmental and social costs. These, in turn, have brought about immense ecological degradation, depletion of resources, and social inequalities, which further compel a strategic realignment toward sustainability within supply chain management. It includes sustainable supply chains: practices that minimize environmental impact, foster social equity, and advance economic viability (Hawon et al., 2023). This helps organizations be in step with business operations motivated by values from society and environmental stewardship.

Sumon et al. (2023), articulated that most of the demands for sustainable supply chains in the USA are driven by the urgent challenge of climate change. The world has been fully informed through repeated calls by the Intergovernmental Panel on Climate Change regarding the urgent need to reduce the level of greenhouse gas emissions to be able to mitigate global warming. As per Rahmnan et al. (2024), supply chains are large contributors to these emissions, as most of the global carbon footprint consists of activities such as transportation, manufacturing, and waste contributed by supply chains. It would be a great avenue for companies to reduce their emission through employing green practices, such as route optimization in transportation, reduction of packaging waste, and transitioning to renewable sources of energy, among others.

**Regulatory and Market Pressures**

Zeeshan et al. (2024), indicated that over the last couple of years, sustainability has increased in its emphasis from regulatory bodies and more environmentally conscious consumers alike. The US government has been in talks for some time to reduce carbon emissions through the imposition of policies, such as the Inflation Reduction Act of 2022, which allows for huge financial incentives for companies to develop and use cleaner technologies. In addition, today's consumer is full of awareness about environmental issues and is willing to pay a greater price for sustainable products. Islam et al. (2023b), upholds that with every passing day, companies are under pressure not only to remain within the ambit of the law but also to increase their reputation and competitiveness in the market. In these scenarios, more sustainable models of supply chains have started to emerge, in which carbon footprint reduction is one of the major motives.

**Resource Efficiency**

According to Khallaf (2024), resource efficiency is one of the major areas in the context of sustainable supply chains in the U.S. Since the population is growing, it will lead to an increase in water and energy consumption, among other resources. Unsustainable extraction and consumption patterns will inevitably threaten the availability of these basic resources for the current generation as well as future generations. Thus, sustainable supply chains promote the conservation of resources through recycling processes, circular economy initiatives, and sourcing practices. By reducing the production of waste and encouraging businesses to use resources more effectively, resilience and profitability will increase, along with contributing to broader economic sustainability.

**Corporate Social Responsibility**

Pal (2023), asserted that Corporate Social responsibility is equally important in the discussion of sustainable supply chains. The ethical impacts of supply chain practices have also received considerable attention, especially regarding labor conditions, human rights, and community impacts. Unsustainable practices pertain to exploitation, poor working conditions, and adverse impacts on local communities, common in developing countries. A commitment to sustainability in supply chains can ensure that an organization accounts for fair labor, diversity, and inclusion, with meaningful community development (Hasan et al., 2023). This will enhance brand reputation and build loyalty among increasingly ethical consumers.

**The Role of AI in Enhancing Decision-Making for Sustainable Supply Chains**

Shobhana (2024), contends that Artificial Intelligence (AI) entails a myriad of technologies, comprising machine learning (ML), deep learning, natural language processing (NLP), and computer vision. These are sets of enabling technologies that allow massive amounts, patterns, and predictions developed on data and drive automation in decision-making processes. Eyo-Udo (2024) posited that AI can be leveraged for operational effectiveness, diminishing generated waste, and lowering $CO_2$-equivalent emissions for supply chain management. AI use in logistics goes beyond the specifics of an area of application to everything from demand forecasting and inventory to route optimization, supplier evaluation, and sustainability reporting. With AI integrated into decision-making in the supply chain, an organization would be better positioned to make data-driven decisions aligned with its sustainability objectives.

**AI-Powered Decision-Making Application in Sustainable Supply Chains**

**Demand Forecasting and Inventory Optimization:**

Accurate demand prediction assists organizations reduce excess inventory, diminishing waste, and the need for additional storage, which frequently requires significant energy consumption. The forecast of demand, if accurate, will help reduce excess inventory. Demand can be predicted even more efficiently by using machine learning algorithms on analysis of historic sales data, seasonality, and external factors such as economic indicators (Chen et al., 2024). This would optimize inventory levels, reducing the carbon footprint associated with overproduction, warehousing, and disposal of unsold goods.

**Route Optimization and Fleet Management:**

Transportation is one of the largest contributors to supply chain emissions in America. AI-driven route optimization algorithms find the most efficient routes that can be taken, which reduces fuel consumption and, subsequently, emissions. Similarly, AI can optimize fleet management to monitor vehicle performance and predict maintenance needs for scheduling deliveries with minimum distances of travel (Shobhana, 2024). For example, UPS uses an AI-powered system called ORION, an abbreviation for On-Road Integrated Optimization and Navigation; this technology saves millions of gallons of fuel annually through efficient routing of deliveries.

**Supplier Evaluations and Risk Management:**

Buiya et al. (2024), articulated that AI helps organizations to analyze data from their suppliers related to environmental impact, compliance, and social responsibility. NLP listens for news, reports, and social media regarding any sort of red flags related to suppliers that will help organizations make more sustainable procurement decisions. This, in turn, helps organizations to reduce overall emissions indirectly by opting to work with those suppliers whose carbon footprint is reduced.

**Smart Warehousing and Energy Efficiency:**

Using automation and intelligent management software systems, AI can extend warehouse optimizing below consumption of energy. With the help of autonomous robots, AS/RS-DDN (Automated Storage and Retrieval Systems), and smart lighting, carbon footprint reduction can be achieved for warehousing activities( Shawon et al, 2024). Energy usage can also be optimized for lower emissions on predictive maintenance and energy management systems driven by AI.

**Product Lifecycle Management:**

AI fervently augments PLAs by reading data from raw material sourcing, production, transportation, and end-of-life disposal. With an invaluable environmental understanding of each step, companies are able to make informed decisions for more environmentally appropriate designs for products and processes (Sumon et al. 2023). Artificial intelligence can also support a circular economy by optimizing reverse logistics associated with returning products for refurbishment or recycling.

**Carbon footprint tracking and reporting**

AI-powered analytics platforms empower organizations to efficiently and effectively measure and report their carbon footprint. The platforms could automate data collection from a variety of sources, analyze emissions information, and render regulatory-compliant sustainability reports. Such transparency engenders greater accountability among the companies concerned and helps carbon-reducing mechanisms (Hasan et al., 2024).

**AI-Enhanced Decision-Making System Framework**

**Step 1: Data Acquisition:** The first step for the AI-enhanced decision-making system is robust data acquisition, which includes data from internal systems such as ERP and CRM, and external data including supplier information, market trends, and environmental data. IoT devices can capture real-time data on the consumption of resources, wastes generated, and emissions across the supply chain (Rahman et al. 2023). High-quality and comprehensive data is very important for training AI models in ensuring accurate predictive analytics.

**Step 2: Incorporating AI Analytics:** Once the information is acquired, it needs interpretation using AI analytics. Machine learning algorithms analyze historical data for patterns, trends, and areas where inefficiency may arise. Predictive analytics forecast demand, optimize inventory levels, and assess the sustainability of suppliers in sustainable chains (Zeeshan et al. 2024). For example, AI can forecast fluctuations in demand, thus allowing organizations to adjust their production schedule and avoid waste. Advanced algorithms can also evaluate the level of carbon footprint coming from different supply chain activities and, hence, guide the business by making decisions aligned with sustainability goals.

**Step 3: Decision-Support Systems-** Subsequently, the insights gained through analytics from AI feed into decision-support systems allowing managers to arrive at informed choices. Such a system can also be used for scenario analysis, as it evaluates the potential impacts of various options on sustainability metrics. For instance, managers could simulate the sustainability impacts of the substitution of suppliers, based on their different carbon equivalents, or on the use of different transportation routes that minimize environmental impact(Islam et al, 2023b). By providing actionable insights in an easily understandable format, the decision-support system assists stakeholders in making decisions on sustainability for operational strategy.

**Step 4: Stakeholder Involvement**- Effective stakeholder engagement stands at the core of an AI-enhanced decision-making system. The suppliers, customers, and logistic partners collaboratively work toward creating a transparent and jointly accountable system regarding sustainability goals. Indeed, AI can facilitate such engagement through building platforms where data may be shared for viewpoint exchange (Hawon et al. 2023). For instance, organizations may integrate dashboards where stakeholders can access real-time views from sustainability metrics, enabling joint initiatives and entrainment of sustainable practices throughout supply chains.

**Step 5**: **Performance Evaluation:** The last part of the framework concerns performance evaluation, where sustainability outcomes would be continuously monitored and assessed. The AI systems can monitor KPIs for carbon emission, resource use, and waste generation. Regular assessment can allow organizations to find out parts that need a change and adapt their strategy in the right direction (Khan et al. 2024). This is an iterative process that builds a culture of continuous improvement, whereby sustainability efforts remain attuned to evolving environmental standards and stakeholder expectations.

**Challenges in the Implementation of AI for Sustainable Chains**

**Data Quality and Integration**

Hasan et al. (2024), reported that one of the pressing challenges in successfully deploying AI in supply chains is ensuring data quality and integration. Inconsistent or incomplete data leads to incorrect predictions and, consequently, poor decisions. This will require an organization to invest in a strong data management system to ensure the AI algorithms access substantial, reliable data.

**Resistance to Change**

Cultural resistance to the use of new technologies is another noteworthy barrier to the implementation of AI solutions. Workers may strongly distrust AI-driven decision-making because they feel threatened by the loss of jobs or lack of control over key processes (Buiya et al., 2024). Organizations should create an innovative culture and train employees to understand the benefits derived from AI.

**Ethical and Regulatory Considerations**

As AI becomes more instrumental in supply chain management, companies in America must navigate regulatory and ethical considerations. Data privacy, algorithmic bias, and transparency of decision-making top the list. Companies will have to balance compliance with regulations and ethical standards in AI practices (Zeeshan et al., 2024).

**High Implementation Costs:**

Implementation of AI solutions is relatively expensive for SMEs: Technology acquisition, training employees, and integration in the prevailing processes have huge costs associated with them. However, the long-term benefits of AI in reducing emissions and improving efficiency can outweigh the initial investment (Rahman et al. 2024).

**Complexity of AI Models:**

The development and deployment of AI models is a specialized art. Many companies do not possess the in-house capability to build and maintain AI-driven systems (Khan et al., 2024). Secondly, AI models can also be arcane and hard to interpret; hence, it is difficult for a decision-maker to trust and act upon AI recommendations.

**Conclusion**

AI has the potential to be a game-changer in the future of how supply chains function, ensuring a much more efficient, cost-effective operation, and sustainable process. Supply chains in the US account for an enormous percentage of all carbon emissions, so decision-making using AI in these types of systems is key to achieving environmental targets. That can be achieved by the optimization of demand forecasting, route planning, supplier evaluation, and energy management. Besides, AI can help reduce carbon footprints and meet regulatory and consumer demands for sustainability. Many challenges stand in the way of the successful adoption of AI in supply chains; poor data quality, high costs of implementation, and resistance from stakeholders are among them. With technologies continuing to evolve, the integration of AI with other emerging technologies, blockchain, and edge computing will further enhance the sustainability of supply chains. The journey to sustainable supply chains is more of a technological challenge but also a strategic imperative. An organization embracing AI-driven sustainability will surely thrive in a fast-paced business evolution while contributing toward a greener and more sustainable future for all.